\documentclass{ws-mpla}
\usepackage[super]{cite}
\usepackage{graphicx}
\newcommand{\ba}{\begin{eqnarray}}
\newcommand{\ea}{\end{eqnarray}}
\newcommand{\be}{\begin{equation}}
\newcommand{\ee}{\end{equation}}
\newcommand{\bz}{\begin{itemize}}
\newcommand{\ez}{\end{itemize}}

\newcommand{\da}{\delta}
\newcommand{\la}{\lambda}
\newcommand{\ka}{\kappa}
\newcommand{\za}{\zeta}
\newcommand{\sa}{\sigma}

\newcommand{\oa}{\omega}
\newcommand{\Ga}{\Gamma}

\newcommand{\La}{\Lambda}

\newcommand{\cF}{{\cal F}}
\newcommand{\cP}{{\cal P}}
\newcommand{\cQ}{{\cal Q}}

\newcommand{\cO}{{\cal O}}
\newcommand{\cW}{{\cal W}}

\newcommand{\w}{\widetilde}
\newcommand{\p}{\partial}

\newcommand{\n}{\nabla}

\newcommand{\ra}{\rightarrow}
\newcommand{\Ra}{\Rightarrow}

\newcommand{\LF}{\left(}
\newcommand{\RF}{\right)}
\newcommand{\LT}{\left[}
\newcommand{\RT}{\right]}
\newcommand{\Ld}{\left.}
\newcommand{\Rd}{\right.}

\newcommand{\kb}{\bar{k}}
\newcommand{\pb}{\bar{p}}
\newcommand{\4}{\frac{1}{4}}

\newcommand{\mx}{\mbox}
\newcommand{\mt}{\mathtt}
\newcommand{\mand}{\mx{ and }}

\newcommand{\where}{\mx{ where }}
\newcommand{\with}{\mx{ with }}

\newcommand{\ie}{{\it i.e.\ }}

\newcommand{\non}{\nonumber\\}
\newcommand{\Fc}{\mathcal{F}}

\newcommand{\pd}{\partial}

\newcommand{\D}{\nabla}

\begin{document}

\markboth{Tirthabir Biswas and Spyridon Talaganis}
{String-inspired infinite-derivative theories of gravity: a brief overview}

%%%%%%%%%%%%%%%%%%%%% Publisher's Area please ignore %%%%%%%%%%%%%%
%\catchline{}{}{}{}{}
%%%%%%%%%%%%%%%%%%%%%%%%%%%%%%%%%%%%%%%%%%%%%%%%%%%%%%%%%%%%%%%%%%%

\title{String-inspired infinite-derivative theories of gravity: A brief overview
}

\author{\footnotesize Tirthabir Biswas}

\address{Department of Physics, Loyola University, 6363 St. Charles Avenue,\\ Campus Box 92, New Orleans, LA 70130, USA\\
{\it tbiswas@loyno.edu}
}

\author{\footnotesize Spyridon Talaganis}

\address{Consortium for Fundamental Physics, Lancaster University, Lancaster LA1 4YB, UK\\
{\it s.talaganis@lancaster.ac.uk}
}

\maketitle

\begin{abstract}
In String Theory, there often appears a rather interesting class of higher-derivative theories containing an infinite set of derivatives in the form of an exponential. These theories may provide a way to tame ultraviolet (UV) divergences without introducing ghost-like states. Here we provide a brief overview on the progress that has been made over the last decade to construct such infinite-derivative theories of gravity (IDG) which may be able to address the singularity problems in gravity. In the process, we will present some general results that apply to covariant torsion-free metric theories of gravity.
\keywords{Space-time singularities; modifications of general relativity; metric perturbations.}
\end{abstract}

%\ccode{PACS Nos.: include PACS Nos.}
%%%%%%%%%%%%%%%%%%%
\section{Introduction}
The theory of General Relativity (GR) has an ultraviolet (UV) problem. At the classical level, this
is manifested in cosmological or black-hole type singularities.  Over the years there have been several attempts to tame these divergences: (a) approaches where space-time metric is replaced by a more ``fundamental'' description (such as via the AdS/CFT correspondence) and the resolution of the singularities is expected to only be evident in this new language. (b) Approaches which do not  alter the GR description of space-time dynamics but tinkers with the matter stress energy tensor sourcing gravity. Either exotic forms of matter sources such as the Casimir~\cite{casimir} or  BCS gap~\cite{bcs}  energy are considered, or one modifies the way ordinary matter contributes to stress energy tensor at the high energy regime~\cite{lqc}. (c) Approaches where at least an approximate space-time continuum description is assumed to hold even at Planckian (or close to Planckian) curvatures but the Einstein's equations are modified at the ultra-violet (UV) scale so as to lead to nonsingular Black hole/Big Bang solutions. This paper is devoted to one such  approach based on a covariant torsion-free metric theory of gravity containing an infinite set of covariant derivatives similar to what one finds in stringy constructions, such as open string field theory~\cite{sft}, p-adic theory~\cite{padic} and strings on random lattices~\cite{random}.
%Towards the end of the paper, we will also discuss how one may go about trying to address the issue of quantum divergences  in the context of these infinite derivative theories of gravity (IDG).

Modifying gravity consistently is a nontrivial task. Firstly, there are   constraints from different tests of GR~\cite{will}. These tests however, mostly probe physics at the infrared (IR) scale, and therefore any UV modification of gravity that is needed to solve the singularities, will typically avoid these experimental bounds. For completeness sake we will enumerate the conditions needed to realize GR at low energies by computing the Newtonian potentials; this is also going to be helpful for those who are working on modified gravity theories to explain dark energy.

Secondly, since any covariant modification of gravity inevitably involves higher derivatives which are notorious for introducing ghosts in the theory~\cite{gravity-ghosts}, one has to be very careful. For instance, in Ref.~\cite{Stelle}, it was
demonstrated that fourth order theories of gravity are renormalizable,
but unfortunately they included ghost states. The presence of ghosts imply that the quantum theory is non-unitary. In an effective theory approach one may be tempted to ignore the problem, but the presence of ghosts almost always signal the presence of  classical instabilities, thus it becomes imperative that we avoid them. Thirdly, one also needs to ensure that tachyonic instabilities are absent to avoid any superluminal propagation which plagues several dark energy motivated modified gravity theories~\cite{gravity-ghosts}. What is rather convenient is that to address all these questions, one mostly needs to look at the behavior of small fluctuations around a relatively small set of physically important classical backgrounds, such as the Minkowski space-time, Freedman-Lemaitre-Robertson-Walker (FLRW) cosmological backgrounds, and spherically symmetric metrics describing spatial regions with astrophysical densities. So far, progress have been made to understand perturbations around the simplest case of the Minkowski background and our main aim in this paper is to provide an overview of the results obtained so far. The final question we will address is whether any consistent gravitational theory can resolve the UV infinities in GR. 

We will see that ghost-free infinite-derivative theories of gravity (IDG)  provides hope for resolving the classical black hole, and big bang singularities possibly by becoming asymptotically free at short distances. We will summarize the main findings so far on the subject, and end with a brief treatise on the UV properties of quantum loops in the IDG framework.

The paper is organized as follows: in section~\ref{sec:classification}, we will derive  the form of the most general gravitational action that is relevant to study fluctuations around the Minkowski background and obtain the linearized field equations around the Minkowski space-time. Next, in section~\ref{sec:stability}, we will derive the propagator for the different degrees of freedom present in the metric and discuss the conditions necessary to avoid ghosts and tachyons. In section~\ref{sec:singularity}, we will address the classical singularities in gravity within the linearized approximation scheme. This will be followed in section~\ref{sec:exact}, by a study of exact bouncing solutions that are known for certain special classes of IDG theories and that may be relevant for Bounce inflation models~\cite{bounce-inflation}. Section~\ref{sec:quantum} will essentially provide a preview of the ongoing work~\cite{scalar-gravity} on understanding quantum properties of IDG theories. We will conclude in section~\ref{sec:conclusions} outlining the important challenges that lie ahead for IDG theories.
%%%%%%%%%%%%%%%%%%%%%%%%%%%%%%%%%%%%%%%%%%%%%%%%%%%%%%%%%%%%%%%%%%%%%
\section{Covariant Theories admitting Minkowski Vaccum}~\label{sec:classification}
%%%%%%%%%%%%%%%%%%%%%%%%%%%%%%%%%%%%%%%%%%%%%%%%%%%%%%%%%%%%%%%%%%%%
\subsection{Relevant Quadratic Action \& Popular Examples}
As we discussed in the introduction, for investigating the theoretical and observational consistency of the gravitational models, for most purposes it is sufficient to only consider  quadratic actions in fluctuations around the background metric. Accordingly, our aim in this section is to arrive at the most general form of the gravitational action that is relevant for studying the classical and quantum properties of the fluctuations around the simplest background metric, that of Minkowski space-time. Incredible as it may sound, it is actually possible to write down the most general form of a covariant metric based gravitational action rather compactly as~\cite{BGKM}
\be
S=\int d^4x\sqrt{-g}\LT \cP_0+\sum_i \cP_i\prod_I (\hat{\cO}_{iI} \cQ_{iI})\RT
\label{general}
\ee
where $\cP,\cQ$'s are composed only of the Riemann and the metric tensor, while the differential operators $\hat{\cO}$'s are made up solely from covariant derivatives, and contains at least one of them. We do not specify anything about the index structure of these quantities.
Since we want the Minkowski space to be a legitimate vacuum solution of the theory, and the Riemann tensor vanishes for the Minkowski space-time, it is clear that $\cP,\cQ$'s must be nonsingular for the Minkowski metric and therefore be expandable as a power series around $R_{\mu\nu\la\sa}=0$. Now, we are interested in looking at metric
fluctuations around the Minkowski background
\be
g_{\mu\nu}=\eta_{\mu\nu}+h_{\mu\nu}\,,
\ee
and consider terms in the action that are quadratic in $h_{\mu\nu}$, and every appearance of the Riemann
tensor contributes an $\cO(h)$ term in the action. Hence,
 only terms that are products of at most  two
curvature terms are relevant, and higher ones simply do not play any role
in this analysis, and can be ignored. Thus up to integration by parts, the most general relevant action  is of the form
\be\label{action1}
S=\int d^4x\sqrt{-g}\LT{R\over 2} + R_{\mu_1\nu_1\la_1\sa_1}
\cO_{\mu_2\nu_2\la_2\sa_2}^{\mu_1\nu_1\la_1\sa_1}
R^{\mu_2\nu_2\la_2\sa_2}\RT,
\ee
where $\cO$ is a differential operator containing covariant derivatives
and $\eta_{\mu\nu}$, and we have set $M_p=1$.

Using the symmetry properties of the Reimann tensor and the Bianchi identities, it turns out that, one can rewrite the most general action (\ref{action1}) in terms of three arbitrary dimensionless functions, ${\cal F}_i(\Box)$'s~\cite{BGKM},
\begin{equation}
S = \int d^4x\ \sqrt{-g}\left[{R\over 2}+R \cF_1\LF\Box\RF R+R_{\mu\nu}
\cF_2\LF\Box\RF R^{\mu\nu}+ R_{\mu\nu\la\sa} \cF_{3}\LF\Box\RF R^{\mu\nu\la\sa} \right]\ .
\label{minkowski}
\end{equation}
Let us now look at some of the well-known gravitational theories.
The Starobinsky model corresponds to the case when $\cF_2=\cF_3=0$ and $\cF_1=f_0>0$. Stelle's 4th order theory corresponds to the case when all the $\cF$'s are non-zero but they are just constants. Another class of gravitational theories that have become rather popular have the form (\ref{general}) but with only the $\cP_0$ term which is a function of the Ricci scalar and the Gauss Bonnet invariant:
\be
\cP_0=\cP(R,G)\with G=R^2-4R_{\mu\nu}R^{\mu\nu}+R_{\mu\nu\la\sa}  R^{\mu\nu\la\sa}\ .
\ee
According to our prior arguments, for the purpose of considering fluctuations up to quadratic order in $h$ around the Minkowski,  only the terms that are linear or quadratic in $R$, and linear in $G$ needs to be considered, all other terms have at least three curvatures. Further since $G$ is a total derivative in four dimensions, we are back to the Starobinsky model.

Finally, let us illustrate how  to obtain the relevant action of the form (\ref{minkowski}) by looking at the 6th order theory of gravity given by the action~\cite{sixth}:
\ba
S &=& \int d^4x\ \sqrt{-g}\left[{R\over 2}+f_{1,0}R^2+f_{2,0}R_{\mu\nu} R^{\mu\nu}+ f_{3,0}R_{\mu\nu\la\sa} R^{\mu\nu\la\sa} \Rd\non
&+& c_1R^3+c_2RR_{\mu\nu} R^{\mu\nu}+c_3RR_{\mu\nu\la\sa} R^{\mu\nu\la\sa}+c_4R_{\mu\nu} R^{\nu\la}R_{\la\sa} g^{\mu\sa}+c_5R_{\mu\nu} R_{\la\sa}R^{\mu\la\nu\sa}\non
&+& c_6R_{\mu\nu} R^{\mu\la\rho\sa}R^{\nu}{}_{\la\rho\sa}+c_7R_{\mu\nu\la\sa} R^{\mu\rho\la\za} R_{\rho}{}^\nu{}_\za{}^\sa\ +c_8R_{\mu\nu\la\sa} R^{\mu\nu\rho\za} R_{\rho\za}{}^{\mu\nu}\non
&+& f_{1,1}R\Box R+f_{2,1}R_{\mu\nu}\Box R^{\mu\nu}+ f_{3,1}R_{\mu\nu\la\sa}\Box R^{\mu\nu\la\sa}\Big]\ .
\label{6th}
\ea
Since all the terms in the second and third line contains three curvatures they do not contribute to the quadratic fluctuations and the equivalent quadratic curvature action for the fluctuations simply reduces to (\ref{minkowski}) with
\be
\cF_i= f_{i,0}+f_{i,1}\Box\ .
\ee
%%%%%%%%%%%%%%%%%%%%
\subsection{Perturbative Field Equations}
Let us now obtain the free quadratic (in $h_{\mu\nu}$) part of the  action (\ref{minkowski}).
Since the curvature vanishes on the Minkowski background, the two
$h$ dependent terms must come from the two curvature terms present.  This
means the covariant derivatives take on their Minkowski values. Many of the terms simplify and combine to eventually produce the  action~\cite{BGKM,BCKM}
\begin{alignat}{5}
& S_q   =  -  \int d^4x\Big[
\frac{1}{2}h_{\mu\nu} a(\Box) \Box h^{\mu\nu}
 +  h_{\mu}^{\sa} b(\Box)\p_{\sa}\p_{\nu}h^{\mu\nu}\label{lin_act} \\ \notag
 & +  h c(\Box)\p_{\mu}\p_{\nu}h^{\mu\nu}
+ \frac{1}{2}h d(\Box) \Box h
 +  h^{\la\sa} \frac{f(\Box)}{\Box}\p_{\sa}\p_{\la}\p_{\mu}\p_{\nu}h^{\mu\nu}\Big]\, ,
\end{alignat}
\begin{eqnarray} \label{abc}
\where\ a(\Box) &=&   1+2 {\cal F}_2 (\Box) \Box +8 {\cal F}_{3} (\Box) \Box \ ,\\
b(\Box) &= & -1-2 {\cal F}_2 (\Box) \Box - 8 {\cal F}_{3} (\Box) \Box \ ,\\
c(\Box) &=& 1-8 {\cal F}_1(\Box) \Box - 2 {\cal F}_2 (\Box) \Box \ ,\\
d(\Box) &=& -1+8{\cal F}_1(\Box) \Box  +2 {\cal F}_2 (\Box) \Box \\
\label{f_term}
\mand f(\Box) &=&   8 {\cal F}_1 (\Box) \Box +4 {\cal F}_2 (\Box)\Box+8  {\cal F}_{3}(\Box) \Box .
\end{eqnarray}
%\end{alignat}
The above can be thought of as a higher derivative generalization of the
action considered in~\cite{peter} where
 $a,b,c,d$ were just constants. The function
$f(\Box)$ appears only in higher derivative theories.

  From the explicit
expressions we observe the following relationships:
\begin{equation}\label{Grelations1}
a +b = 0; ~~~
c +d = 0;~~~
b+c + f = 0 \, .
\end{equation}
So, we are left with only two independent arbitrary functions. The field equations can be written in the form
\begin{eqnarray}\label{linearized-eqn}
& &a(\Box) \Box h_{\mu\nu}   +   b(\Box)\p_{\sa}( \p_{\nu}h_{\mu}^{\sa} + \p_{\mu}h_{\nu}^{\sa})
 +  c(\Box)(\eta_{\mu\nu}\p_{\rho}\p_{\sa}h^{\rho\sa}  +\p_{\mu}\p_{\nu}h)\non
& +&   \eta_{\mu\nu}d(\Box)\Box h
 +    f(\Box) \Box^{-1} \p_{\sa}\p_{\la}\p_{\mu}\p_{\nu}h^{\la\sa}=-2\ka\tau_{\mu\nu}\with \ka=M_p^{-2}
\end{eqnarray}
{\it This is the most general form of the field equations for linear fluctuations around the Minkowski metric resulting from an arbitrary covariant metric theory of gravity.}

A few comments are now in order: While the matter sector obeys stress energy conservation,  the geometric part
is also conserved as a consequence of the generalized Bianchi identities:
\ba
\nabla_\mu \tau^{\mu}_\nu = 0 & = & (a+b)\Box h^\mu_{\nu,\mu} + (c+d)\Box \p_\nu h+(b+c+f )h^{\alpha\beta}_{,\alpha\beta\nu}\, .
\ea
It is now clear why (\ref{Grelations1}) had to be satisfied. What is also reassuring is that these same conditions not only ensure that the different spin degrees of the metric (a spin 2, a vector and two scalars) decouple but they also eliminate the vector and one of the scalars which are typically ghost like. Finally, let us note that as long as the $\cF$'s are analytic functions around $\Box=0$ we have from (\ref{abc})
\be
a(0)=c(0)=-b(0)=-d(0)  =1\mand f(0)=0\ ,
\label{GRlimit}
\ee
and this ensures that we recover GR in the infrared limit. This also means that
$\lim_{\Box\ra 0} f(\Box)\propto\Box\ ,$
 and therefore   $f(\Box)/\Box$ remains analytic around $\Box= 0$. $a(\Box)$ and $c(\Box)$ are also obviously analytic as long as $\cF$'s are. Conversely~\footnote{There is one other possibility where the $\cF$'s could have poles or other singularities at $\Box=0$ in such a way that the singularities cancel in $a(\Box)$ and $c(\Box)$. This could be interesting for dark energy phenomenology. In fact, modified gravity theories where even $a(\Box)$ has a pole such a $\sim {1/ \Box}$ have been considered as dark energy models which aim to modify the behavior of gravity at cosmic distances, for a recent review see~\cite{woodard}.}, the requirement that we recover GR in the far IR implies that  the $\cF$'s, and hence, $a,c$, are smooth functions around $\Box=0$.
%%%%%%%%%%%%%%%%%%%%%%%%%%%%%%%%%%%%%%%%%%%
\section{Stability \& Consistency of Nonlocal Theories around Minkowski Background}~\label{sec:stability}
\subsection{Propagator and physical poles}
%%%%%%%%%%%%%%%%%%%%%%
 We are now well-equipped to calculate
the propagator. The above field equations can be written in the form
\be
\Pi_{\mu\nu}^{-1}{}^{\la\sa}h_{\la\sa}=\ka\tau_{\mu\nu}
\ee
where $\Pi_{\mu\nu}^{-1}{}^{\la\sa}$ is the inverse propagator. One can obtain
the propagator using the spin projection operators $\{P^2,P_{s}^0,P_{w}^0,P_m^1\}$,
corresponding to the spin-2, the two scalars, and the vector
 degree of freedom respectively, please see~\cite{peter,BGKM} for details.
 Considering each sector separately and
taking into account the constraints (\ref{Grelations1}),
one eventually arrives at a rather simple result~\cite{BGKM}
\be\label{prop}
\Pi={P^2\over ak^2}+{P_{s}^0\over (a-3c)k^2}\, .
\ee
We observe that the vector multiplet and the $w$-scalar have disappeared as foreshadowed, and the
remaining $s$-scalar has decoupled from the tensorial structure. Further, courtesy (\ref{GRlimit})
 as $k^2\ra 0$, we have only the physical graviton propagator:
\ba
\lim_{k^2\ra 0}\Pi^{\mu\nu}{}_{\la\sa}&=&{P^2\over k^2}-{P_{s}^0\over 2 k^2}\,.
\ea
Although, the $P_s$
residue at $k^2=0$ is negative, it is a benign ghost. In fact,  $P_s^0$ has
precisely the coefficient to cancel the unphysical longitudinal degree of freedom
in the spin two part~\cite{peter}.
%%%%%%%%%%%%%%%%%%%%%%%%%%%
\subsection{Consistency Conditions on $a(\Box),c(\Box)$}
Let us next look at some well known special cases. Arguably, the so-called $F(R)$
 gravity theories have been the most well studied, where $F(R)={R\over 2}+f_{1,0}R^2+\dots$ In this case, only the $\cF_1$ appears as a higher derivative contribution.
 Since $a=1$, it is easy to see that only the
$s$-multiplet propagator is modified.  It now has two poles:
$\Pi\sim -P_s^0/2k^2(k^2+m^2)$, with $m^2=1/12f_{1,0}$. The  $k^2=0$ pole has, as usual,  the wrong sign
of the residue, while the second pole has the correct sign.  This represents an
additional scalar degree of freedom coming from the presence of the $R^2$ correction. Equivalently, this theory can be recast into a Brans-Dicke type non-minimally coupled scalar-tensor theory of gravity. Since we want the scalar degree to be stable, this means its mass must be non-tachyonic:
\be
m^2>0\Ra f_{1,0}>0
\ee
While $F(R)$ theories are ghost-free and have been widely studied as  inflationary and dark energy models they cannot resolve the classical singularities of GR, see for instance~\cite{BMS} for a detailed argument.

A theory which is known to be able to resolve classical singularities~\cite{robert-bh} (and in fact even tame quantum divergences~\cite{Stelle}) is the so called Fourth order modification involving $R_{\mu\nu}R^{\mu\nu}$. This corresponds to having a $\cF_2$ term which modifies the spin-2 propagator: $\Pi\sim P_2/k^2(k^2+m^2)+\dots$. The second pole necessarily has the wrong residue sign and therefore unfortunately contains the infamous Weyl ghost~\cite{peter,gravity-ghosts}.

In fact, the above situation is quite typical: $F(R)$ type models can be ghost-free,  but they do not improve UV behavior, while modifications involving $ R_{\mu\nu\la\sa}$'s can improve the UV behavior but typically they contain the Weyl ghost! To see how we may be able to overcome this problem let us summarize the conditions necessary to have a stable and ghost-free theory around the Minkowski background:
\bz
\item Imposing that we recover GR at large distances imply that $a(\Box),c(\Box)$ must be analytic around $\Box=0$.
\item The condition that there be no Weyl ghost tells us that $a(\Box)$ cannot have any zeroes.
\item The fact that the scalar mode doesn't have any ghosts, imply that $a-3c$ can at most have one zero. This, in turn, means that we should be able to express $c(\Box)$ in the form
\be
c(\Box) = {a(\Box) \over 3}\left[1+ 2\LF1-\frac{\Box}{m^2}\RF\tilde{c}(\Box)  \right],
\ee
where $\tilde{c}$ cannot have any zeroes, and must also be analytic around $\Box=0$.
\item Further, the fact that we do not want any tachyonic modes, imply $m^2>0$. One could however let $m^2\ra\infty$.
\ez
%%%%%%%%%%%%%%%%%%%%%%%%%%%
\subsection{General Categories}
From the above conditions on $a(\Box),c(\Box)$ we can now deduce the kind of gravitational theories that can admit a consistent Minkowski vacuum:
\begin{itemize}
\item $F(R,G)$ theories with the additional condition
\be
\Ld{\p^2 F\over \p R^2}\right|_{R=G=0}\geq 0
\ee
to ensure that the scalar mode, if present, is non-tachyonic.
\item Higher Curvature Theories: Modifications involving more than two curvature terms do not suffer from problems of ghosts around Minkowski as they simply do not contribute to the free part of the gravitational action, and thus remain completely unconstrained!
\item Nonlocal theories: These are theories when $a,\w{c}$ do not contain any zeroes in the complex plane but contain poles, or other singularities (except at $\Box=0$). A typical example would be $a(\Box)=c(\Box)=(\Box -m^2)^{-1}$.
Whether such theories can be free from instabilities is not obvious and requires further study. They clearly make the theory even more divergent in the UV and therefore cannot address the UV problems.
\item IDG theories: These are ``mildly'' non-local gravity theories which contains an infinite set of derivatives in the form of entire functions. In the rest of the manuscript we will mainly focus on this third category, although some of the results quoted are more general. In terms of the degrees of freedom, we can further differentiate between different possibilities:
    \begin{itemize}
     \item[Type G:] If $m^2\ra \infty$ and $\w{c}=1$, the theory contains only the {\bf G}raviton as a propagating degree of freedom, and the entire function $a=c$ controls the UV modification.
     \item[Type P:] If $a=1$, $m^2\ra \infty$ and  $\w{c}$ is an entire function, then the theory resembles a {\bf P}-adic type theory coupled non-minimally to gravity. So, there are no new perturbative degrees of freedom, but the scalar contains nonlocal interactions.
     \item[Type S:] If $a=1$, $m^2$ is finite and $\w{c}$ is an entire function, then we have a {\bf S}FT (String Field Theory)-type theory which is non-minimally coupled to ordinary GR.
     \item[Type B:] Both $a$ and $\w{c}$ are nontrivial entire functions. In this case {\bf B}oth the scalar and the graviton are nonlocal. Depending upon whether $m^2$ is finite or not, the scalar will be of the SFT or $p$-adic type respectively.
     \end{itemize}
\end{itemize}
%%%%%%%%%%%%%%%%%%%%%%%%%%%%%%%%%%%%%%%
\section{Classical Singularities}\label{sec:singularity}
%%%%%%%%%%%%%%%%%%%%%%%%%%%%%%%%%%%%%%%%
In the rest of this paper we are going to focus on the IDG theories and try to assess whether they can cure the UV problems that plague GR. In this section we will look at  the black hole and the big bang singularities and try to make progress staying within the linearized approximation.
\subsection{Asymptotic freedom \& the status of the Black hole singularity}
%%%%%%%%%%%%%%%%%%%%%%%%%%%%%%%%%%%%%%%%%%%%%%%%%%%%%%%%%%%%%%%%%%%%%%%%%%%
To illustrate how nonlocal modifications can help, let us look at the G-type models with $f=0$ or equivalently $a=c$. The propagator then simplifies to:
\be\label{prop1}
\Pi^{\mu\nu}{}_{\la\sa}={1\over k^2a(-k^2)}\LF P^2-\2P_{s}^0\RF .
\ee
We are left with only a single arbitrary function $a(\Box)$ which only modifies the graviton propagator without introducing any new states as compared to GR.

Let us analyze the two scalar Newtonian potentials , $\Phi(r),~\Psi(r)$, corresponding to the metric
\be
ds^2=-(1+2\Phi)dt^2+(1-2\Psi)dx^2\with |\Phi|,|\Psi|\ll 1\ .
\label{newtonian-metric}
\ee
 As is usual, we want to
solve the linearized modified Einstein's equations (\ref{linearized-eqn})
%, which for our simplified case reads
%\begin{eqnarray}\label{G-eqn}
%a(\Box) [\Box h_{\mu\nu}  -   \p_{\sa}\p_{(\nu}h_{\mu)}^{\sa} +  \eta_{\mu\nu}\p_{\rho}\p_{\sa}h^{\rho\sa}  +\p_{\mu}\p_{\nu}h
%-   \eta_{\mu\nu}\Box h]=\ka\tau_{\mu\nu}\ ,
%\end{eqnarray}
for a point source:
\be
\tau_{\mu\nu}=\rho\da_\mu^0\da_\nu^0=m\da^3(\vec{r})\da_\mu^0\da_\nu^0\ .
\ee
Due to the Bianchi identities we only need to solve the
trace and the $00$ component of (\ref{linearized-eqn}), which for a static metric, simplifies to
\ba
(a -3c)\Box h+(4c -2a+f)\p_{\mu}\p_{\nu}h^{\mu\nu}
 &=& 2\ka \rho \nonumber \\
a\Box h_{00} + c\Box h  -c  \p_{\mu}\p_{\nu}h^{\mu\nu}& =&  -2\ka\rho\,.
\label{trace-00}
\ea
In terms of the Newtonian potentials we get
\ba
(a -3c)[\n^2 \Phi-2\n^2 \Psi]
 &=& \ka \rho \nonumber \\
(c-a)\n^2\Phi -2 c\n^2\Psi  & =&  -\ka\rho\,.
\label{trace-00a}
\ea
Specializing to type G case we have
\ba
2a(\n^2)\n^2\Phi =2 a(\n^2)\n^2\Psi  =  \ka\rho=\ka m\da^3(\vec{r}).
\label{newtonian}
\ea
where let us remind ourselves that  $a(\Box)$ must
be an entire function.  Let us, in fact, consider the simplest entire function which is just an exponential,
\be
a(\Box)=e^{-\Box/M^2}\ ,
\label{exponential}
\ee
that appears so ubiquitously in String theory~\cite{sft,padic,random}.

Taking the Fourier components of (\ref{newtonian}) one straightforwardly obtains
\be
\Phi(r) = -\ka m\int {d^3p\over (2\pi)^3}\ {e^{i\vec{p}\vec{r}}\over 2 p^2 a(-p^2)}= -{\ka m\over 4 \pi^2 r}\int {dp\over p} \frac{\sin{p\, r}}{a(-p^2)}.
\ee
We note that the $1/r$ divergent piece comes from the usual GR action, but now it
is ameliorated. For (\ref{exponential}) we have
\be
\Phi(r) = - {\ka m\over 4\pi^2 r}\int {dp\over p} e^{-{p^2\over M^2}} \sin{(p\,r)}=-{\frac{\ka m}{8\pi r}}\mt{erf}\left(\frac{r M}{2} \right),
\ee
and the same for $\Psi(r)$.  We observe that as $r \ra \infty$, $\mt{erf}( r)\ra 1$, and
we recover the GR limit. On the other hand, as $r\ra 0$, $\mt{erf}(r)\ra r$, making the
Newtonian potential converge to a constant $ \sim mM/M_p^2$. Although, the matter source has
a delta function singularity, the Newtonian potentials remain finite! Further, provided
\be
mM\ll M_p^2\ ,
\label{mass-condition}
\ee
 our linear approximation can be trusted all the way to $r\ra 0$.

Firstly, we see that an UV modification involving the entire function is able to realize ``asymptotic freedom'' where the force vanishes as one approaches $r\ra 0$. This turns out to be true for a large class of entire functions. Note that any entire function can be written as
\be
a(\Box)=e^{-\xi(\Box)}\ ,
\ee
where $\xi$ is an analytic function. It is then easy to see that for any polynomial $\xi$, as long as the highest power has a positive coefficient, we will have an asymptotically free theory.
Secondly, it is clear that for small masses our theory provides a very different description of spacetime as compared to GR. In fact, according to our model there are no  black-hole like solutions (no horizon and no singularity) as long as the source mass satisfies (\ref{mass-condition}). Our analysis, unfortunately, cannot say anything about large mass black holes because the Newtonian potentials become too large for us to be able to trust the perturbative calculations. Understanding large mass black hole geometry therefore remains an outstanding challenge for the IDG theories.
%%%%%%%%%%%%%%%%%%%%%%%%%%%%%%%%%%
\subsection{Big Bang Singularity}
%%%%%%%%%%%%%%%%%%%%%%%%%%%%%%%%%%%
To  study the Big Bang singularity problem let us consider an FLRW type metric in conformal time,
\be
ds^2=a^2(\tau)\eta_{\mu\nu}dx^{\mu}dx^{\nu}\with a(\tau)= 1+ \da a(\tau)\ ,\ \da a\ll 1\ ,
\ee
and a diagonal stress energy tensor given by
\be
\tau_{00}=\rho\ ;\ \tau_{ii}=p\with p=\oa \rho\mand -1\leq\oa\leq 1\ .
\label{ideal-fluid}
\ee
Essentially this means that we only need to consider the trace part of the metric:
\be
h_{\mu\nu}=\4 \eta_{\mu\nu} h\approx 2\eta_{\mu\nu}\da a(\tau)
\label{FLRW}
\ee
Substituting (\ref{FLRW}) in (\ref{linearized-eqn}) one finds
\begin{eqnarray}
-{1\over 8}[a(\Box)-3c(\Box)][\eta_{\mu\nu} \Box-\p_{\mu}\p_{\nu}]h=\ka\tau_{\mu\nu}\non
\end{eqnarray}

For the 00 component it is easy to see that the left hand side vanishes, meaning that solutions can only be obtained for vanishing stress energy tensor~\footnote{For an ideal fluid satisfying (\ref{ideal-fluid}) a vanishing energy density implies vanishing pressure.}. Furthermore, from the $ii$ component one finds
$$
[a(-\p_{\tau}^2)-3c(-\p_{\tau}^2)]\ddot{h}=0
$$
According to~\cite{ic}, this has two solutions. Either
\be
\ddot{h}=0\ , \mx{ or }\ [a(-\p_{\tau}^2)-3c(-\p_{\tau}^2)]h=0
\label{lin-cos}
\ee
In the former case we have the solution, $h=\tau/\tau_0$. This is a monotonically expanding solution and unfortunately the linearized approximation breaks down when the scale factor approaches the big bang singularity: $a\approx 1+h/8$, so as $a\ra 0$, $|h|\ra 8>1$. Thus we cannot say anything about what happens near the singularity.

The latter possibility only has solutions~\cite{ic} iff the function $a(\Box)-3c(\Box)$ has a zero which is also precisely the condition for the theories to have an extra degree of freedom! Thus it seems that one cannot address the  cosmological singularity problem at the linearized level in G-type theories. The situation is the same with ordinary GR, and the problem arises because around the Minkowski background the terms that are linear in $h_{\mu\nu}$ vanish for all G-type theories. All this really means is that to gain insight into cosmological solutions one has to go beyond the free theory  and include cubic interactions in the action. Since, a general expression for the cubic interactions have not yet been derived, we will now specialize to S-type theories where an additional scalar degree of freedom is present, \ie when $a\neq c$, and we have a solution to the equation
\be
a(m^2)-3c(m^2) =0\with m^2>0
\ee
In this case we have a cosmological solution given by
\be
(\p_t^2+m^2)h=0\Ra h(t)=h_0\sin(mt)
\ee
As long as $h_0\ll 1$ our linearized approximation is valid and the metric describes a nonsingular space-time with small oscillations around the Minkowski metric:
\be
a(t)=1+8h_0\sin(mt)\ .
\ee
The above result is encouraging but it is only valid in the absence of any stress-energy tensor. The natural question is whether one can obtain similar nonsingular solutions with  non-zero energy densities. As discussed before, if one had the cubic interaction terms for the nonlocal actions, one could continue the analysis perturbatively, but here we will focus on a different approach based on finding exact solutions to the gravitational field equations by imposing additional constraints.
%%%%%%%%%%%%%%%%%%%%%%%%%%%%%%%%%%%%%%%%%%%%%%%%%%%%%%%%%%%%%%%%%%%%%%%%%
\section{Exact Bouncing Solutions: Background \& Perturbations}\label{sec:exact}
%%%%%%%%%%%%%%%%%%%%%%%%%
Over the last decade or so, IDG theories have gained in interest due to its various novel cosmological applications involving bouncing cosmologies, see for instance~\cite{BMS,BKM,KV,BKMV,BigBang}, accelerating cosmologies (inflation and the current cosmic speed-up), see for instance~\cite{accelerating}, as well as some other applications, see for instance~\cite{others}. Here, we will focus on the possible resolution of the Big Bang/Crunch singularity.
%%%%%%%%%%%%%%%%%%%%%%%%%%%%%%%%%%%%%%%%%%%%%%%%%%%%%%%%%%%%%
\subsection{Bouncing Backgrounds}
In Refs.~\cite{BMS,BKM,KV} the cosmology for the special class of ghost-free actions with $\cF_2=\cF_3=0$, and a non-zero
\be
\cF_1=\cF=\sum_{n=0} f_n\Box^n\ ,
\ee
was investigated in the presence of a cosmological constant, $\La$, and radiation. For this case we have
\be
a=1\mand c= 1-8\cF(\Box)\Box
\ee
This is a P or S-type  theory depending on whether the combination,
\be
a-3c=-2[1-12\cF(\Box)\Box]\ ,
\ee
has no or single root respectively. All the nonlocality resides on the scalar sector while the graviton propagator remains unchanged. The Einstein-Schmidt (ES) field equations for such theories read~\cite{Schmidt,BMS}
\ba
&[&1+4\Fc(\Box)R]G^\mu_\nu
-\sum_{n=1}
^\infty
f_n\sum_{l=0}^{n-1}\Bigl[g^{\mu\rho}\pd_\rho\Box^l  R  \pd_\nu\Box^{n-l-1}  R
+g^{\mu\rho}\pd_\nu\Box^l  R  \pd_\rho\Box^{n-l-1}  R  {}\non
&+&\delta^\mu_{\nu}\left(
\pd^\rho\Box^l  R  \pd_\rho\Box^{n-l-1}  R  +\Box^l  R  \Box^{n-l}  R
\right)\Bigr]+ R \Fc(\Box) R\delta^\mu_{\nu}-4(\D^\mu\pd_\nu-\delta^\mu_{\nu}
\Box)\Fc(\Box) R=-\Lambda \delta^\mu_{\nu}+{T}^\mu_\nu\non
\label{eqEinsteinRonlyupdown}
\ea
where ${T}^\mu_\nu$ is the  energy--momentum tensor of
radiation and
\begin{equation}
G^\mu_\nu=R^\mu_\nu-\frac12\delta^\mu_\nu R\ ,
\end{equation}
is the Einstein tensor. The above equation has a rather simple looking trace equation:
\begin{equation}
%\begin{split}
R- 2\sum_{n=1}^\infty
f_n\sum_{l=0}^{n-1}\Bigl(\pd_\mu\Box^l  R  \pd^\mu\Box^{n-l-1}  R
+2\Box^l  R  \Box^{n-l}  R\Bigr)-  12\Box\Fc(\Box)R=4\Lambda\, ,
%\end{split}
\label{eqEinsteinRonlytrace}
\end{equation}
Note, that radiation being conformal does not contribute to the trace equation.

This trace equation can be solved by imposing  an additional constraint~\cite{BMS,BKM,KV}
\begin{equation}
\Box R-r_1R-r_2=0\with r_1\neq 0 \ .
 \label{ansatz}
\end{equation}
The differential equations simply reduce to a set of algebraic equations
\ba
\Fc'{(r_1)}&=&0\\
 r_2&=&{}-\frac{r_1[1-12\Fc(r_1)r_1]}{4[\Fc(r_1)-f_0]}\\
\Lambda &=&{}-\frac{r_2}{4r_1}. \label{r2lambda}
\label{r1}
\ea
We note that by virtue of the Bianchi identities, solving the trace of the ES equations
essentially guarantee that we have a solution for the complete ES
equations provided we include appropriate radiation density:
\begin{equation}
\rho=\rho_0\left(\frac{a_0}{a}\right)^4\ .
\label{matter13}
\end{equation}
In the cosmological context  the ansatz (\ref{ansatz}) becomes a third order differential equation for the Hubble parameter, $H=\dot a/a$,
\begin{equation}
\label{EquH}
 \dddot H+7H\ddot H+4\dot H^2+12H^2\dot H
 ={}-2r_1H^2-r_1\dot H-\frac{r_2}{6}\ ,
\end{equation}
where dot defines derivative w.r.t physical time. The solutions to the above equation, their stability, and asymptotic behaviours have been studied in details in~\cite{BKM}. It was found that several non-singular solutions exist where the universe bounces from a phase of contraction to a phase of expansion, thus providing a resolution to the big bang singularity problem. By looking at the ``$00$'' component of the ES equations, one can calculate the radiation energy density at the bounce point, $\rho_0$, which must be positive:
\begin{equation}
\rho_0=\frac{3(r_1-4 f_0 r_2)(r_2-12h_1^2)}{12r_1^2-4r_2}
\label{FRW_eqEinsteinRonlyansatztt}
\end{equation}
where $h_1=\dot{H}$ characterizes the acceleration of the universe at the bounce point and plays the role of an ``initial condition''.

We see that within the context of the ansatz (\ref{ansatz}),  the  solutions are characterized by three independent parameters, $r_1$, $ f_0$, and $ \cF_1$ which fixes a specific value of the cosmological constant via (\ref{r2lambda}).
There are three cosmological scenarios one can envisage:
\begin{itemize}
\item $\La<0,\ r_1>0\Ra r_2>0$: One obtains a cyclic universe scenario where the universe undergoes successive bounces and turnarounds. Qualitatively speaking the bounce is caused by the higher derivative modifications, while the turnarounds occur when the radiation energy density cancels the negative cosmological constant. Such scenarios can give rise to ``cyclic inflationary'' models if one is able to incorporate entropy production~\cite{CI}.
\item $\La>0,\ r_1<0\Ra r_2>0$: One obtains a bouncing universe scenario where a phase of contraction gives way to a phase of super-inflating expansion. An exact example is the Ruzmaikin-type solution
which incidentally contains no radiation~\cite{KV}. Unfortunately, in these scenarios the universe is stuck in a quantum gravitational phase accelerating faster and faster, and we never recover a GR phase. Clearly such solutions are not phenomenologically viable and hence we disregard these scenarios.
\item $\La>0,\ r_1>0\Ra r_2<0$: This represents a geodesic
completion of an inflationary space-time via a non-singular bounce. It is easy to see that the field equations (\ref{eqEinsteinRonlyupdown}) without  matter admits constant curvature vacuum solutions, the de
Sitter (dS) solution. The dS solutions is in fact
exactly the same as that of GR since for
constant $R$, all the higher derivative terms in the field equations
vanish, and one just obtains from ~(\ref{eqEinsteinRonlytrace})
\begin{equation}
\label{de SitterSol}
R_{dS}={4\frac{\Lambda}{M_P^2}}.
\end{equation}
The space-time for this case asymptotes to the dS solution as  radiation dilutes away very quickly during the cosmological constant dominated phase.

One can actually find an exact solution of this type given by
\begin{equation}
a(t) = a_0\cosh{\sqrt{\frac{r_1}{2}}t},\quad\Rightarrow \quad H=\sqrt{\frac{r_1}{2}}\tanh\left(\sqrt{\frac{r_1}{2}}t\right)
\label{FRW_Hexact}
\end{equation}
 This solution satisfies the ansatz with the specific parameter combination
\begin{equation*}
\Box R = r_1 R-6r_1^2.
\end{equation*}
Substituting $r_2=-6 r_1^2$ into (\ref{r2lambda}) and (\ref{FRW_eqEinsteinRonlyansatztt}) we get
\begin{equation}
\label{lambdaMp}
\Lambda=\frac{3}{2}r_1M_P^2 \quad\mand\quad \rho_0=-\frac{27}{2}\la\Fc_1r_1^2\Ra \Fc(r_1)<0
\end{equation}
for radiation to be non-ghost like.
This last constraint turns out to be a general constraint which must be satisfied in order to realize geodesically complete de Sitter space-time, but it is relatively easy to find ghost-free $\cF$'s satisfying all the necessary criteria to obtain a nonsingular bounce~\cite{BMS,BKM}.
\end{itemize}
%%%%%%%%%%%%%%%%%%%%%%%%%%%%%%%%%%%%%%%%%%%%%%%%%%%%%%%%%%%%%%%%%
\subsection{Bounce Inflation: Perturbative stability}
Bounce inflation~\cite{Bounce-inflation,BI-anomalies,BI-flucts} has become a particularly interesting cosmological scenario for several reasons: It provides a nonsingular geodesic completeness to inflationary space-time; it is naturally suited to address the low multipole anomalies~\cite{Bounce-inflation} in CMB as the near exponential inflation is necessarily preceded by a super inflationary phase;  it also provides a way of stretching matter and gravity wave fluctuations that may be produced during the contracting phase to cosmological scales during  subsequent inflation thereby making them relevant for CMB~\cite{BI-flucts}. Since our IDG models can provide such Bounce inflationary solutions, it becomes interesting to investigate the viability of these  models further by looking into the fluctuations around the backgrounds. Such an investigation has proved to be rather challenging and technical but progress have been made~\cite{BKMV}. Here we summarize the main results:
\begin{itemize}
\item In the late time $dS$ limit there always exists four modes: the first two modes are similar to what you expect in GR coupled to  radiative fluctuations. The fluctuations decay as the universe inflates. The two other modes only decay if, and only if, $r_1>0$ and therefore to have a stable inflationary background we must impose this condition.
\item Additionally, there may be more solutions depending upon whether the function $\cW(\Box)$ has roots, $\{w_i^2\}$, or not, where
\begin{equation}
\cW(\Box)=3\cF(\Box)+(R_{dS}+3r_1)\frac{\cF(\Box)-\cF_1}{\Box-r_1}\,.
\label{deltatraceGIconst}
\end{equation}
For real $\oa_i^2$, we can only get a stable deSitter solution if $\oa_i^2\geq 0$~\footnote{$\oa_i^2\approx 0$, is a rather interesting and special case corresponding to the presence of a massless degree of freedom. In this case $\Phi-\Psi$ tends to a constant at late times, rather than vanishing as in GR. In fact, if one assumes the usual Bunch-Davis type subHubble vacuum initial conditions for this field, then one will end up with a scale invariant spectrum for these modes. This is very similar to the curvaton scenario that is routinely considered in inflationary cosmology, and deserves further study.}. If $\oa_i^2$ is complex, then stability of the late time dS is ensured only if the real part of $\nu$ given by,
\be
\nu^2=\LF{\frac94}-\frac{\oa_i^2}{H^2}\RF\ ,
\ee
is less than $3/2$.
\item Studying the perturbations near the bounce is more difficult. However,since we have already provided the criteria to ensure that none of the fluctuations grow during inflation, all we need to check is that there are no modes which become singular at any finite time. The study in~\cite{BKMV} showed that the perturbations seem to pass through smoothly around the bounce. This gives more credibility to the viability of the bounce inflation scenario within the IDG framework.
\end{itemize}
%%%%%%%%%%%%%%%%%%%%%%%%%%%%%%%%%
\section{Quantum Divergences}\label{sec:quantum}
%%%%%%%%%%%%%%%%%%%%%%%%%%%%%%%
In this paper we have seen how IDG theories may provide a way to consistently address classical singularities present in GR. Can these theories be touted as full fledged quantum gravity theories, or are we only supposed to consider them as effective theories of gravity valid only up to a certain energy scale? The quantum calculations in infinite-derivative scalar theory analogues have actually provided considerable success in understanding stringy phenomena such as emergence of massive stringy excitations~\cite{string-states}, Regge behavior~\cite{regge} and stringy thermodynamics~\cite{thermo}. Phenomenological quantum field theory models have also been constructed~\cite{Moffat,BKR,okada}. One of the main feature of these models is that the exponential suppression of the propagator  makes Feynman integrals converge in the UV. So, can these UV finiteness be exploited to obtain a renormalizable quantum theory of gravity? Unfortunately, in gauge theories (and gravity is one) the propagators and interactions are invariably related and the IDG vertices contain exponential enhancement factors. This makes addressing the issue of renormalizability in any theory of gravity, including IDG theories, a rather challenging task. In this paper, we are going to preview some of the work that has been done in this regard~\cite{scalar-gravity} in a toy scalar model ``mimicking'' gravity. For other efforts to construct a quantum IDG theory, see~\cite{Tomboulis,Moffat-gravity,Krasnov,BG,Modesto,Anslemi}.
%%%%%%%%%%%%%%%%%%%%%%%%%%%%%%%%%%%%%%%%%%%%%%%%%%%%%%%%%%%
\subsection{Divergence Structure in Gravitational Theories}
Let us start by quickly reviewing the counting of the superficial degree of divergence (the power, $D$, of growth with cut-off scale) of a Feynman graph in GR:
\be
D = 4 L - 2 I + 2 V,
\ee
where $L$ is the number of loops, $V$ is the number of vertices and $I$ is the number of internal propagators. Using the following topological relation,
\be
\label{eq:topological}
L = 1 + I - V \mx{ we find } D = 2 L + 2.
\ee
Thus, the superficial degree of divergence increases as the number of loops increases making the theory non-renormalizable.

In the IDG theory, we've got a modified superficial degree of divergence since one now needs to count the exponent~\cite{Modesto}. Since the exponential contributions of the propagators and the vertex factors cancel each other,  using \eqref{eq:topological} we have
\be
\label{eq:E}
E = V - I = 1 - L.
\ee
Thus, if $L \geq 2$, $E$ is negative and the loop amplitudes are superficially convergent. In contrast with the power-counting argument in GR, it thus seems that the IDG theories may be renormalizable!
%%%%%%%%%%%%%%%%%%%%%%%%%%%%%%%%%%%%%%%%%%%%%%%%%%%%
\subsection{Quantum Loops in a Toy Model of Gravity}
Ideally, we would like to obtain the interaction terms for IDG theories and then use it to compute the Feynman diagrams. This however turns out to be an extremely hard task both in terms of computing the interactions~\footnote{Gravity contains interactions of all orders. Since these interactions are suppressed by Planck scale, the cubic interactions are the most relevant, but even obtaining the complete cubic vertex proves to be a highly nontrivial task.} as well as using the complicated Feynman vertex functions to compute the Feynman diagrams. Thus, in order  to  understand whether non-localization can at all help to tame the UV divergences in gravitational theories, in~\cite{scalar-gravity} the UV properties of a toy scalar field theory has been studied. The scalar theory preserves a combination of the conformal and  shift symmetries (like gravity around Minkowski space-time),  $\phi \ra (1+\epsilon)\phi + \epsilon$. This ensures that the most relevant property of gravitational theories when it comes to studying quantum UV properties, \ie the compensatory nature of the exponential suppressions and enhancements in the propagators and  vertex factors respectively, is retained.
The action containing cubic interactions is given by
\be
\label{eq:action}
S = \frac{1}{2}\int \mathrm{d}^4 x \, \LF  \phi \Box a(\Box) \phi\RF + \frac{1}{M_p} \int \mathrm{d} ^ 4 x \, \LF \4 \phi \partial _ {\mu} \phi \partial ^ {\mu} \phi + \4 \phi \Box \phi a(\Box) \phi - \4 \phi \partial _ {\mu} \phi a(\Box)  \partial ^ {\mu} \phi \RF\ .
\ee

The propagator in momentum space is given by
\be
\Pi (k ^ 2)= \frac{- i}{k^2 e ^ {\kb ^ 2}},
\ee
 while the vertex factor $V(k_{1},k_{2},k_{3})$ is given by
\be
\label{eq:V}
V (k _ {1}, k _ {2}, k _ {3}) = \frac{i}{4} \LF k_{1}^2+k_{2}^2+k_3^2 \RF \LT 1 -  e ^ {\kb _ {1} ^ {2}} -  e ^ {\kb _ {2} ^ {2}} - e ^ {\kb _ {3} ^ {2}}\RT \Ra V(k)\equiv V (k, -k, 0) = - i k ^ {2} e ^ {\kb ^ {2}}\ .
\ee
The barred  momenta denote that the momenta have been divided by the mass scale $M$.
The $1$-loop, $N$-point function with zero external momenta~\footnote{It should be noted that the symmetry factor is equal to $2N$ when $1$PI corrections to the effective potential are considered (the external points are not fixed in that case). When computing a Green's function, the symmetry factor is equal to $2$ for $N=1,2$ and to $1$ for $N>2$.} evaluates to
\be
\label{eq:N}
\Ga_N = \frac{(-1) ^ {N}}{2N} \frac{i \Lambda ^4}{32 M _ {p} ^ {N} \pi ^ 2}\ ,
\ee
where $\La$ is a hard cutoff. The power of divergence doesn't change with $N$ and is exactly as one would expect from the counting argument (\ref{eq:E}).

If we evaluate the two-loop Feynman diagrams for zero external momenta, we again find a $\La^4$ divergence as would be expected since the 2-loop diagrams contain $1$-loop subdivergences. Again this suggests  that we do not get any additional divergences as we proceed from $1$-loop to $2$-loops, corroborating (\ref{eq:E}).  However, this does not guarantee renormalizability. To prove renormalizability we must make sure that after we eliminate the 1-loop divergences (by possibly adding suitable counterterms), the remaining external momentum dependence of the 1-loop diagrams is, at least, milder than the bare vertex in the UV to make the higher loops convergent. For a more detailed argument see~\cite{scalar-gravity}. With this in mind let us check the external momentum dependence of the $1$-loop, $2$-point function:
\be
\Ga_{2,1}(p^2) = \frac{i}{2 i ^ {2} M _ {p} ^ {2}} \int \frac{\mathrm{d} ^ 4 k}{(2 \pi) ^ {4}} \, \frac{V ^ {2} (-p, \frac{p}{2} + k, \frac{p}{2} - k)}{(\frac{p}{2} + k) ^ {2} (\frac{p}{2} - k) ^ {2} e ^ {\LF\frac{\pb}{2} + \kb\RF ^ {2}} e ^ {\LF\frac{\pb}{2} - \kb\RF ^ {2}}}\,\ .
\label{2pt-ext}
\ee
In dimensional regularization scheme we obtain an $\epsilon ^ {-1}$ pole, as expected, which can be eliminated using suitable counterterms~\cite{scalar-gravity}. The problem however is that as $p\ra \infty$, $\Ga_{2,1}(p^2) $ now diverges exponentially as $e^{\frac{3\pb^2}{2}}$! This, for instance, would make the $2$-loop diagram, which has a renormalized $1$-loop 2-pt subdiagram, divergent. So, is all hope of renormalizability lost?

It is well known that a $1$-loop, $2$-point insertion in any graph can be replaced by a sequence of such insertions separated by the bare propagator. The sequence of all the graphs gives rise to a geometric series which can be summed in the appropriate regime and then analytically continued to the entire complex $p^2$-plane giving rise to the ``dressed'' propagator,
\be
\w{\Pi}(p^2) = \frac{\Pi(p^2)}{1-\Pi(p^2)\Ga_{2,1\mt{PI}}(p^2)} \ .
\ee
Remarkably, now the fact that at large momentum, $\Ga_2$ grows even more strongly than $e^{\pb^2}\sim \Pi^{-1}$, makes the dressed propagators  more exponentially suppressed than their bare counterparts: $ \w{\Pi}(p^2)\ra  e^{-\frac{3\bar{p}^2}{2}} $.
In particular, this makes all $1$-loop graphs, higher than 2-point,  UV-finite! The UV part of the $2$-loop integrals become finite too once the 1-loop 2-pt contribution is renormalized.

To summarize, while a more rigorous investigation involving higher loops is in order, the initial results from quantum loop computations in IDG-type theories seem very promising. In particular, it seems that these theories may have a way of dealing with the compensating nature of the exponential enhancements and suppressions present in vertices and propagators in gravitational theories.
%%%%%%%%%%%%%%%%%%%%%%%%%%%%%%%%%%%%%%%%%%%
\section{Conclusion}\label{sec:conclusions}
%%%%%%%%%%%%%%%%%%%%%%%%%%%%%%%%%%%%%%%%%%%
Over the last decade, work done by several groups have yielded rather promising signs for the infinite-derivative theories of gravity. We have seen that they can be free from instabilities and ghosts around Minkowski background, but at the same time resolve black hole and big bang singularities. This is in contrast with finite higher derivative theories which have been found to be either ghost-free or singularity-free, but not both. We have also started to understand the quantum properties of these type of theories and address the question of renormalizability. There are several outstanding challenges that however remain, and let us conclude by stating some of these:
\bz
\item To conduct a perturbative study of the big bang singularity problem for generic matter sources, we need to go beyond the linearized theory and include cubic interactions. These terms are not known presently.
\item The exact cosmological solutions were only obtained in the presence of a cosmological constant. A realistic cosmological scenario must include a graceful exit from the inflationary phase. At the moment a scheme to analyze such a transition is lacking and any progress in this direction would be greatly beneficial.
\item While exact nonsingular cosmological solutions have been found for IDG theories, it has only been possible to address the black hole singularity in the linearized limit valid for small masses. Understanding the spatial geometry around astrophysical masses  remains a very interesting and important future task.
\item Unlike traditional field theories where the vacuum is often unique around which perturbative analysis needs to be carried out, one must deal with several different background metrics in gravitational theories. While issues of stability and consistency has been addressed around the Minkowski space-time, one needs to urgently extend the analysis to the dS and FLRW space times. This may end up providing us with additional constraints on higher curvature terms (analysis around Minkowski only cares about terms containing two curvatures), and suggest a way to extend the IDG theories further.
\item Several of the loop integrals are computed in Euclidean space, and with the help of prescriptions such as analytic continuation and Cauchy Principle Value theorem. While unitarity and causality of infinite-derivative theories have been formally argued in previous literature~\cite{unitarity}, one perhaps needs to explicitly check that the various prescriptions used to calculate the loop integrals do not spoil unitarity.
\item As we have briefly seen, quantum UV behavior of IDG theories seem promising, but so far calculations have only been carried out up to 2 loops. We need to investigate how the UV convergence properties extend to higher loops. Eventually, one would also obviously like to move from the simple toy model to actual IDG theories.
\ez
In brief, IDG theories have shown enough of a promise to warrant further investigations.
%%%%%%%%%%%%%%%%%%%%%%%%%%
\section*{Acknowledgments}
TB would like to thank Jose Cembranos, Joseph Kapusta, Tomi Koivisto, Alex Koshelev, Anupam Mazumdar,  Nobuchika Okada, and Warren Siegel, for several fruitful collaborations in the general field of nonlocal field theories, as well as numerous discussions on the subject. TB would also like to thank Robert Brandenberger, Leonardo Modesto and John Moffat for engaging exchanges. ST is supported by a scholarship from the Onassis Foundation.
%\section*{References}

\end{document}